\documentstyle[preprint,revtex]{aps}
\math-with-secnums
\begin{document}
\draft
\begin{title}
{\bf Theory of coupled $\pi$--trinucleon  systems}
\end{title}
\author{L. Canton and G. Cattapan}
\begin{instit}
Istituto Nazionale di Fisica Nucleare e
Dipartimento di Fisica dell' Universit\'a\\
via Marzolo 8, Padova I-35131, Italia
\end{instit}
\receipt{December 15, 1993}

\begin{abstract}
We derive the dynamical equations which consistently couple
the four--body ($\pi$NNN) system
to the underlying three--nucleon system.
Our treatment can be considered the proper generalization
of the Afnan--Blankleider equations for the coupled NN--$\pi$NN
system. The resulting connected--kernel equation resembles in structure
the Yakubovskii--Grassberger--Sandhas equation for the standard
four--body problem, but involves 24 chain--labelled components
(rather than the usual 18 ones)
and allows for a consistent evaluation of reaction amplitudes
involving $\pi$ absorption/production.
Preprint no. DFPD 93/TH/77
\end{abstract}

\pacs{PACS numbers: 25.80Ls, 21.45.+v, 11.80.Jy, and 13.75.Gx}

\narrowtext

Consider a picture where three nucleons interact among themselves
through two--body (NN) potentials and with the additional pion
through another two--body ($\pi$N) potential;
then the problem involves (only!) the complexities of a conventional
4--body problem. It can be completely (and, unambiguously) solved
by considering connected--kernel equations such as those arising from
the Grassberger--Sandhas--Yakubovskii formalism\cite{grassberger67}.
However, even if this solution would have
been pursued thoroughly (and with the related outstanding numerical
efforts!) our understanding of the system and related processes would be
rather limited. This because a fundamental aspect -- the
absorption process -- is still
missing in the theory. This channel couples the original four--body system
with the corresponding three--nucleon system.
Clearly, the argument applies as well if we start with a pure
three--nucleon system because, above the pion threshold, the
(inverse) pion production process couples the three--nucleon space to
the one with the additional pion.

In the past decades, much attention has been devoted to the treatment
of simple systems involving baryons and pions. The situation has been
recently reviewed in Ref.\cite{phillips93}; for a
discussion of related conceptual difficulties see also \cite{sauer85}.

The NN--$\pi$NN system has been studied using different
diagrammatic approaches by Thomas, Rinat, Afnan, and Blankleider
\cite{thomas79,afnan80}, and by Avishai
and Mizutani~\cite{avishai83}. Formally, identical equations hold
for the NNN--$\pi$NNN problem\cite{avishai83',cattapan93},
but in this case the disconnectedness of the associated kernel
makes the problem ill--conditioned, in that unphysical (spurious)
solutions join up with the correct one. To our knowledge, there has
been only one work \cite{avishai83'} making substantial improvements on
this crucial difficulty. In their approach, the authors relied
substantially on the isobar picture, which allowed them
to recast the original NNN--$\pi$NNN problem into an effective
multichannel problem coupling isobars (also called, in other contexts,
correlated pairs, or ``quasiparticles") with the underlying
three--nucleon space. Connectedness was achieved
first in the ``standard" four--body sector using the conventional
Grassberger--Sandhas scheme, and then introducing later the additional
complexities due to pion absorption/production,
{\it via} a formal separation based
on a two--potential scattering formula.
By the admission of the authors themselves,
their final equations are extremely involved, and their approach in fact
demands for ``more complete and aesthetic descriptions".
We think we have matched here both these requirements.
In fact, we have obtained connected--kernel equations without using the isobar
(quasiparticle) representation; rather, we used the genuine chain--labelled
formalism of the four--body theory and generalized it to include the
effects of pion absorption/production. This allowed us to treat
consistently and
{\it at the same level} the diagrams involving standard
two--body potentials as well as
those involving meson--baryon vertices. The resulting closed set of
coupled equations yields the univocal, exact, and complete
solution of the problem \cite{note1}.

We define the notation.
With $T_{(1|1)}$ we represent the fully unclusterized {\it scattering}
amplitude 4$\leftarrow$ 4 for the $\pi$NNN system.
In the fully unclusterized {\it absorption} amplitude,
denoted as $T_{(0|1)}$,
the initial channel is similarly made by the four unbound particles, while
in the final channel there are three free nucleons, only (without the pion).
The quantities $T_{(1|0)}$ and $T_{(0|0)}$ are the corresponding
similar amplitudes for fully unclusterized $\pi$ production and
three--nucleon scattering, respectively.

It is most convenient to express
these physical fully unclusterized
amplitudes in terms of the solutions of the
Afnan--Blankleider (AB) equations

\begin{eqnarray}
T_{(1|1)}&=&\sum_a t_a +\sum_{ab} t_aG_0U_{ab}G_0t_b,
\label{T}\nonumber\\
T_{(0|1)}&=&\sum_{a} U^\dagger_{a}G_0t_a,\nonumber\\
T_{(1|0)}&=&\sum_{a} t_aG_0U_{a},\nonumber\\
T_{(0|0)}&=& U.
\nonumber\end{eqnarray}

The auxiliary amplitudes $U_{ab}, U^\dagger_a, U_a, U$
satisfy the AB coupled equations
\begin{eqnarray}
U_{ab}&=&G_0^{-1}\bar\delta_{ab}+\sum_c\bar\delta_{ac}t_cG_0U_{cb}
+F_a g_0 U^\dagger_b,
\label{AB}\nonumber\\
U^\dagger_a&=&F^\dagger_a+{\cal V}g_0U^\dagger_a+\sum_c F^\dagger_c
G_0t_cG_0U_{cb},
\nonumber\\
U_a&=&F_a+\sum_c\bar\delta_{ac}t_cG_0U_c+F_ag_0U,
\nonumber\\
U&=&{\cal V}+{\cal V}g_0U+\sum_c{F^\dagger}_cG_0t_cG_0U_c .
\nonumber\end{eqnarray}
Here, the dressed $\pi$NN vertex referring to the $i$--th nucleon,
$f(i)$, has been recast into a new vertex operator
labelled by the three--cluster index $a$,
namely
\begin{equation}
F_a=\sum_{i=1}^{3}\bar\delta_{ia} f(i),\ \ \ \ \
F^\dagger_a=\sum_{i=1}^{3}\bar\delta_{ia} f^\dagger(i),
\nonumber\end{equation}
for production and absorption, respectively.
Because of their importance, these positions deserve further comments.
The index $a$, being the three--cluster label of the $\pi$--trinucleon
sector, denotes also all the pairs one can make out of four bodies.
The index $i$ labels each one of the three nucleons; however, it can also
be used to denote the specific pair ($\pi N_i$). Thus, if $a=(NN)$,
$F_a$ is given by the sum of {\it all} the $\pi NN$ vertices;
if $a=(\pi N_i)\equiv i$, the vertex $\pi N_i N_i$ has to be excluded in
the sum\cite{note2}.

These equations couple together the fully unclusterized amplitude
for the 3$\leftarrow$ 3 process {\it without pions}
with all the  3--cluster amplitudes $U_{ab}$ one can have
{\it with the pion}. In other words, the set of equations for
$U_{ab}, U^\dagger_a, U_a, U$ couple together
the amplitudes for {\it all} ({\it i.e.}, with or without the pion!)
the partitions into three clusters
one can have out of a system made by one pion and three nucleons.
The number of such partitions is 7.

For two nucleons the AB scheme is fully
connected and
it is possible to evaluate all the amplitudes without ambiguities.

With three nucleons we formally rewrite the AB equation
as a {\it super}--Lippmann--Schwinger equation for matrices defined
in {\it all} the partitions in three clusters of the system
\begin{equation}
{\sf T}^{(3)} ={\sf V}^{(3)} +
{\sf V}^{(3)} {\sf G}_{0}^{(3)} {\sf T}^{(3)} ,
\label{(N-1)-LS}
\end{equation}
with the definitions
\begin{eqnarray}
{{\sf G}_{0}}^{(3)}
\equiv&
\left|
\begin{array}{cc}
{G_{0}} ^{(3)}_{(a|b)} & {G_{0}} ^{(3)}_{(a|0)} \cr
 {G_{0}} ^{(3)}_{(0|b)} & {G_{0}} ^{(3)}_{(0|0)} \cr
\end{array} \right|
&=
\left| \begin{array}{ccc}
G_0t_aG_0\delta_{ab} &\ & 0 \\
0 &\ & g_0 \\
\end{array} \right|,
\label{prop}\\
{{\sf V}}^{(3)}
\equiv & \left|
\begin{array}{cc}
{V} ^{(3)}_{(a|b)} & {V} ^{(3)}_{(a|0)} \cr
{V} ^{(3)}_{(0|b)} & {V} ^{(3)}_{(0|0)} \cr
\end{array} \right|
&=
\left| \begin{array}{ccc}
{G_0}^{-1}\bar\delta_{ab} &\ & F_a\\
F^\dagger_b &\ &{\cal V} \\
\end{array} \right|,
\\
{{\sf T}}^{(3)}
\equiv
& \left|
\begin{array}{cc}
{T} ^{(3)}_{(a|b)} & {T} ^{(3)}_{(a|0)} \cr
{T} ^{(3)}_{(0|b)} & {T} ^{(3)}_{(0|0)} \cr
\end{array} \right|
&=
\left| \begin{array}{ccc}
U_{ab} &\ & U_a\\
U^\dagger_b &\ & U \\
\end{array} \right|.
\end{eqnarray}

Then, we introduce in the system
the partitions into {\it two} clusters.
In the 4--body sector, we denote these partitions by $a'$,
while in the 3--nucleon sector, we use the label $a_1$.
For $a'$ we distinguish three classes.
{Type I} denotes the cluster partition ($\pi$NN)N.
{Type II} denotes the two pairs ($\pi$N)(NN).
{Type III} represents the single (NNN)$\pi$ partition.

{}From the standard 4--body theory, it is obvious \cite{sandhas80}
that
\FL
\begin{eqnarray}
V^{(3)}_{(a|b)} \equiv
{G_0}^{-1}\bar\delta_{ab}
&=&
\sum_{a'}
({G_0}^{-1}\bar\delta_{ab}\delta_{a,b\subset a'})
\equiv
\sum_{a'} (v^{(3)}_{a'})_{ab}
\label{sumrule} \\
{\cal V} &=&\sum_{a_1} {\cal V}_{a_1}
\label{sumruleb} \end{eqnarray}
for the separated 4--body and 3--nucleon sectors, respectively.

It is an important observation to realize that a similar
2--cluster sum rule holds also for the corresponding vertices:
\begin{eqnarray}
F_{a}=\sum_{a'} (f_{a'})_{a} \ \ \ \ \ \
F^{\dagger}_{a}=\sum_{a'} (f^{\dagger}_{a'})_{a} ,
\label{sumrulec}\end{eqnarray}
with the internal vertex operator defined as
\begin{equation}
(f_{a'})_{a} ={\sum_{i=1}^{3}}\bar\delta_{ia} f(i)
 \ \ \ \ \ \
{\rm with}\ \ \ \ \  {i,a\subset a'},
\nonumber\end{equation}
and similarly for $(f^\dagger_{a'})_{a}$. This is proved very easily once it
has been observed that two different pairs ($a,i$) unambiguously
identify one single partition $a'$ into two clusters that contains
both of them.

We may also add that, if the pairs ($a,i$) have no
particle in common (including the pion) then $a'$ is of type I,
while it is of type II otherwise.
Therefore there is an internal vertex, and hence a coupling to the
absorption channel, only for partitions of type I and II.
This has very important consequences because
only such two types of partitions unambiguously
identify one single nucleon--nucleon pair $a_1$ in the 3--nucleon sector.

For each of the 2--cluster partitions
we have to introduce the corresponding subamplitudes.

({\bf Type III}). Here, since the vertex operators are not effective,
we have standard 4--body--like sub--amplitudes.
The III--type sub--amplitudes are well--defined through
the AGS (Alt--Grassberger--Sandhas) equation
\begin{eqnarray}
(u_{a'})_{ab}&=&G_0^{-1}\bar\delta_{ab}
+\sum_c\bar\delta_{ac}t_cG_0(u_{a'})_{cb},
\label{III-AB)}\end{eqnarray}
with ${a,b,c\subset a'}$.

({\bf Type I}). Because of the coupling with the 3--nucleon space,
we have not here a standard AGS equation. Rather, we have an
AB equation with corresponding sub--amplitudes
$(u_{a'})_{ab}, (u^\dagger_{a'})_{a}$ and $(u_{a'})_{a},
(u_{a'})_{}$. These equations couple together all the subamplitudes
for the 3--cluster partitions {\it including} the absorption sector
when partitioned in three single nucleons.
The pair $a_1$ is unambiguously determined,
since it corresponds to the unique nucleon--nucleon pair internal to the
given I--type partition $a'$. Thus, we get the I--type subamplitudes
from the connected--kernel coupled equations
\FL\begin{eqnarray}
(u_{a'})_{ab}&=&G_0^{-1}\bar\delta_{ab}
+\sum_c\bar\delta_{ac}t_cG_0(u_{a'})_{cb}
+(f_{a'})_a g_0 (u^\dagger_{a'})_{b},
\nonumber\\
(u^\dagger_{a'})_{a}&=&(f_{a'}^\dagger)_a+{\cal V}_{a_1} g_0
(u^\dagger_{a'})_{a}+\sum_c (f_{a'}^\dagger)_c G_0t_cG_0(u_{a'})_{cb},
\nonumber\\
(u_{a'})_{a}&=&(f_{a'})_a+\sum_c\bar\delta_{ac}t_cG_0
(u_{a'})_{c}+(f_{a'})_ag_0(u_{a'})_{},
\nonumber\\
(u_{a'})_{}&=&{\cal V}_{a_1}+{\cal V}_{a_1}g_0
(u_{a'})_{}+\sum_c({f^\dagger}_{a'})_cG_0t_cG_0(u_{a'})_{c},
\nonumber\\ \label{I-AB} \end{eqnarray}
with ${a,b,c, a_1\subset a'}$.

({\bf Type II}). Like the I--type ones, these
subamplitudes couples the 4--body sector with the 3--nucleon one.
However, the equations for the II--type subamplitudes are
somehow different.
In fact, there is a delicate double--counting
problem connected with the introduction of such sub--amplitudes,
which requires a nontrivial modification with respect to the above
I--type equations. Let's observe the sum rule Eq.~\ref{sumruleb}
for the 3--nucleon sector.
This sum rule has been completely exhausted since it has been
already used for the AB equation for the I--type subamplitudes, and
hence, the equations for the II--type subamplitudes must take into
account this fact:
\FL\begin{eqnarray}
(u_{a'})_{ab}&=&G_0^{-1}\bar\delta_{ab}
+\sum_c\bar\delta_{ac}t_cG_0(u_{a'})_{cb}
+(f_{a'})_a g_0 (u^\dagger_{a'})_{b},
\nonumber\\
(u^\dagger_{a'})_{a}&=&(f_{a'}^\dagger)_a+
\sum_c (f_{a'}^\dagger)_cG_0t_cG_0(u_{a'})_{cb},
\nonumber\\
(u_{a'})_{a}&=&(f_{a'})_a+\sum_c\bar\delta_{ac}t_cG_0
(u_{a'})_{c}+(f_{a'})_ag_0(u_{a'})_{},
\nonumber\\
(u_{a'})_{}&=&\sum_c({f^\dagger}_{a'})_cG_0t_cG_0(u_{a'})_{c} .
\label{II-AB}\end{eqnarray}
Obviously, ${a,b,c\subset a'}$, and ${a_1\subset a'}$ also.

For any type (I, II, and III) of 3--cluster partitions
of the 4--body sector,
the above equations can be formally recast into
a {\it super}--Lippmann--Schwinger matrix equation,
namely
\begin{equation}
{\sf t}^{(3)}_{a'}
={\sf v}^{(3)}_{a'}
+{\sf v}^{(3)}_{a'}
{\sf G}_0^{(3)}
{\sf t}^{(3)}_{a'},
\label{(N-1)-subLS}\end{equation}
with ${\sf G}_0^{(3)}$
defined in Eq.~\ref{prop}.

Since the new T--matrices ${\sf t}^{(3)}_{a'}$ are defined
in the space of all the 3--cluster partitions,
their definitions depend on whether ${a'}$ is of type I and II,
or III. In fact, we have
\begin{eqnarray}
{{\sf t}_{a'}}^{(3)}
&\equiv \left|
\begin{array}{cc}
{t_{a'}} ^{(3)}_{(a|b)} & {t_{a'}} ^{(3)}_{(a|0)} \cr
 {t_{a'}} ^{(3)}_{(0|b)} &{t_{a'}}^{(3)}_{(0|0)}  \cr
\end{array} \right| = \left| \begin{array}{cc}
(u_{a'})_{ab} & (u_{a'})_{a}\\
({u^\dagger}_{a'})_{b}& u_{a'} \\
\end{array} \right|,
\nonumber\end{eqnarray}
\begin{eqnarray}
{{\sf t}_{a'}}^{(3)}
\equiv \left|
\begin{array}{c}
{t_{a'}} ^{(3)}_{(a|b)} \cr
\end{array} \right| =
\left| \begin{array}{c}
{(u_{a'})}_{ab} \\
\end{array} \right|,
\nonumber\end{eqnarray}
for partitions of type (I, II) or (III), respectively.

For III--type partitions the identification of
${\sf v}_{a'}^{(3)}$ corresponds to the standard
4--body choice in the 4--body sector,
\begin{eqnarray}
{{\sf v}_{a'}}^{(3)}
\equiv \left|
\begin{array}{c}
{v_{a'}} ^{(3)}_{(a|b)} \cr
\end{array} \right| =
\left| \begin{array}{c}
{G_0}^{-1}\bar\delta_{ab}\delta_{a,b\subset a'} \\
\end{array} \right|,
\nonumber\end{eqnarray}
because there is no coupling with the 3--nucleon sector.
For I--type partitions, ${\sf v}_{a'}^{(3)}$
has 4 non null sectors
\begin{eqnarray}
{{\sf v}_{a'}}^{(3)}
\equiv \left|
\begin{array}{cc}
{v_{a'}} ^{(3)}_{(a|b)} & {v_{a'}} ^{(3)}_{(a|0)} \cr
 {v_{a'}} ^{(3)}_{(0|b)} & {v_{a'}}^{(3)}_{(0|0)} \cr
\end{array} \right| =
\left| \begin{array}{cc}
{G_0}^{-1}\bar\delta_{ab}\delta_{a,b\subset a'} &
(f_{a'})_{a}\\
({f^\dagger}_{a'})_{b}&{\cal V}_{a_1} \\
\end{array} \right|,
\nonumber\end{eqnarray}
and,
for II--type partitions, ${\sf v}_{a'}^{(3)}$ is defined
according to Eq.~\ref{II-AB},
\begin{eqnarray}
{{\sf v}_{a'}}^{(3)}
&\equiv \left|
\begin{array}{cc}
{v_{a'}} ^{(3)}_{(a|b)} & {v_{a'}} ^{(3)}_{(a|0)} \cr
 {v_{a'}} ^{(3)}_{(0|b)} &{v_{a'}}^{(3)}_{(0|0)}  \cr
\end{array} \right| = \left| \begin{array}{cc}
{G_0}^{-1}\bar\delta_{ab}\delta_{a,b\subset a'} &
(f_{a'})_{a}\\
({f^\dagger}_{a'})_{b}&0 \\
\end{array} \right|.
\nonumber\end{eqnarray}
Thus, all the previous sum rules (Eqs.~\ref{sumrule}, \ref{sumruleb},
and \ref{sumrulec})
can be formally replaced by one single
{\it super} sum rule
\begin{equation}
{\sf V}^{(3)}=\sum_{a'}{\sf v}^{(3)}_{a'}.
\label{supersumrule}\end{equation}

Eqs.~\ref{(N-1)-LS}, \ref{(N-1)-subLS}, and \ref{supersumrule}
are all the basic ingredients we need to derive a new connected--kernel
equation. The solution of such an equation is then used to calculate
the amplitudes ${\sf T}^{(3)}$. On the contrary, a straightforward
solution of the original AB equations would lead to non
univocal results since that kernel is not connected.

The fundamental ansatz
\begin{equation}
{{\sf T}^{(3)} }=\sum_{a'}
{{\sf t}^{(3)} _{a'}} +\sum_{a'b'}
{{\sf t}^{(3)}_{a'}}
{{\sf G}_0}^{(3)}
{{\sf U}^{(3)}_{a'b'}}
{{\sf G}_0}^{(3)}
{{\sf t}^{(3)}_{b'}},
\label{fundamental}
\end{equation}
allows one to write the AB amplitudes in terms of new unknowns
${\sf U}^{(3)}_{a'b'}$. Clearly, Eq.~\ref{fundamental} is an implicit
definition of the new unknowns.

For each pair ($a'$ $b'$), ${\sf U}^{(3)}_{a'b'}$ is a matrix
spanning the 3--cluster partitions, and thus the number of such
matrices is as large as the square of 7.
However, we can also view the set of unknowns
${\sf U}^{(3)}_{a'b'}$ as one {\it single}
matrix defined in the chain space labelled
by one partition in two clusters and one 3--cluster
partition {\it internal} to the given 2--cluster partition. Contrary to
what happens in the
standard 4--body theory, however, here the absorption channel has to be
explicitly included. Thus, there are here
24 components rather that the
18 Yakubovskii ones of the standard four--body problem.
The occurrence of six extra components
is explained by the fact that six different 2--cluster partitions
(namely, all the I-- and II--type partitions) have one additional
component for the absorption channel.

The strategy is to find a new equation for the
unknowns ${\sf U}^{(3)}_{a'b'}$ and then use the ansatz
Eq.~\ref{fundamental} to obtain the AB amplitudes ${\sf T}^{(3)}$.

To get the new equation for ${\sf U}^{(3)}_{a'b'}$,
we substitute Eq.~\ref{fundamental} into the original AB equation
Eq.\ref{(N-1)-LS}, and use for ${\sf V}^{(3)}$ the
sum rule Eq.~\ref{supersumrule}. Then, we take into account
Eq.~\ref{(N-1)-subLS} for all the various types of subamplitudes.
In so doing, we get an equation for ${\sf U}^{(3)}_{a'b'}$
involving sums over all the two--cluster partitions ${a',b'}$.
If we make an identification {\it term by term} of the resulting
equation, we obtain
\begin{equation}
{{\sf U}^{(3)}_{a'b'}} =\bar\delta_{a'b'}{{\sf G}_0^{(3)}}^{-1}+
\sum_{c'}\bar\delta_{a'c'}{\sf t}^{(3)}_{c'}
{{\sf G}_0^{(3)}} {{\sf U}^{(3)}_{c'b'}}.
\nonumber\end{equation}

This equation is our final result, and it is important to observe
that it has formally the same structure of a standard
four--body equation, written in the Yakubovskii--Grassberger--Sandhas
formulation. However, the nature of this equation is different, in that
it has six extra components (24 instead of 18) and couples together
two sectors with different number of particles. We have verified that
the kernel of such an equation is connected. More precisely,
we have verified that two sectors (out of four) are connected already
after one iteration of the kernel, whereas the remaining two sectors
(namely, $\pi$--scattering and $\pi$--production) require a second
iteration.

Also the above equation can be recast into a formal LS equation:
\begin{equation}
{\sf T}^{(2)} ={\sf V}^{(2)} +
{\sf V}^{(2)} {\sf G}_{0}^{(2)} {\sf T}^{(2)}.
\label{(N-2)-LS}
\end{equation}
This LS equation operates at an higher hierarchic level, namely
in the standard chain--labelled space for the 4--body sector,
and in a new space labelled by hybrid chains
for the 3--nucleon sector.
The two spaces are spanned by the chain indices $(a' a)$ [$a\subset a'$],
and $(a',a_1)$ [$a_1\subset a'$], respectively.  The
hybrid nature of the second chain is obvious, since $a'$ is a
2--cluster partition in the 4--body sector, and $a_1$ is another
2--cluster partition in the 3--nucleon sector.

At this level of the hierarchy,
``potentials", ``Green's functions", and ``T matrices'' are defined as
\begin{eqnarray}
{{\sf G}_{0}}^{(2)}
\equiv&
\left|
\begin{array}{cc}
{G_{0}} ^{(2)}_{(a'a|b'b)} & {G_{0}} ^{(2)}_{(a'a|b'b_1)} \cr
 {G_{0}} ^{(2)}_{(a'a_1|b'b)} & {G_{0}} ^{(2)}_{(a'a_1|b'b_1)} \cr
\end{array} \right|
&=
\left| \begin{array}{ccc}
G_0t_aG_0(u_{a'})_{ab}G_0t_bG_0\delta_{a'b'} &\
&
G_0t_aG_0(u_{a'})_{a}g_0\delta_{a'b'} \\
g_0(u^\dagger_{b'})_{b} G_0t_bG_0\delta_{a'b'}
&\ & g_0 (u_{a'}) g_0 \delta_{a'b'}\\
\end{array} \right|,
\nonumber\\
{{\sf V}}^{(2)}
\equiv & \left|
\begin{array}{cc}
{V} ^{(2)}_{(a'a|b'b)} & {V} ^{(2)}_{(a'a|b'b_1)} \cr
{V} ^{(2)}_{(a'a_1|b'b)} & {V} ^{(2)}_{(a'a_1|b'b_1)} \cr
\end{array} \right|
&=
\left| \begin{array}{ccc}
{(G_0t_aG_0)}^{-1}\delta_{ab}\bar\delta_{a'b'} &\ & 0\\
0 &\ &g_0^{-1}\bar\delta_{a'b'} \\
\end{array} \right|,
\nonumber\\
{{\sf T}}^{(2)}
\equiv
& \left|
\begin{array}{cc}
{T} ^{(2)}_{(a'a|b'b)} & {T} ^{(2)}_{(a'a|b'b_1)} \cr
{T} ^{(2)}_{(a'a_1|b'b)} & {T} ^{(2)}_{(a'a_1|b'b_1)} \cr
\end{array} \right|
&=
\left| \begin{array}{ccc}
U_{a'ab'b} &\ & U_{a'ab'b_1}\\
U^\dagger_{a'a_1b'b} &\ & U_{a'a_1b'b_1} \\
\end{array}
\right|.
\nonumber\end{eqnarray}

It must be clearly understood
that {\it when $a'$ and $b'$ are of type {\rm III}
all the hybrid--chain components are absent}, whereas these hybrid--chain
components are substantial when $a'$, $b'$ are of type I and II.
As stated before, for any given I-- or II--type partition $a'$, there is one
and only one $a_1$ partition in the 3--nucleon sector; hence,
$\delta_{a'b'}$ $\Rightarrow$ $\delta_{a_1b_1}$, as well as
$\bar\delta_{a_1b_1}$ $\Rightarrow$ $\bar\delta_{a'b'}$.
Another useful identity is
$\bar\delta_{a'b'}$ $=$ $\bar\delta_{a'b'}\delta_{a_1b_1}$
$+$ $\bar\delta_{a_1b_1}$.

By application of the residue method, it is possible to obtain
amplitudes referring to more clusterized processes.
For instance, the amplitude for the reaction $^3{\rm He}$($\pi$,NN)N
is given by the expression
\begin{eqnarray}
\sum_{a'} <\phi_0|{\sf t}^{(3)}_{a'} {\sf G}^{(3)}_0
{\sf U}^{(3)}_{a'b'}|\tilde\Phi_{b'}>
=&
\sum_{a',a, b}
<\phi_0|(u^\dagger_{a'})_a G_0t_aG_0U_{a'ab'b}|\Phi_{b';b}>
\nonumber\\&+\sum_{a', b}
<\phi_0| (u_{a'}) g_0 U_{a'a_1b'b}|\Phi_{b';b}>.
\nonumber\end{eqnarray}
Here, the sum over $a'$ ranges only over types I and II,
while $b'$ is of type III, and $|\tilde\Phi_{b'}>$ is the
column vector denoting the 3--nucleon bound state.
With varying $b$, $|\Phi_{b';b}>$ are the Faddeev
components  of the target bound state, while $<\phi_0|$ is the
3--nucleon plane--wave state.

In conclusion, we have obtained a connected--kernel scheme
for the coupled NNN--$\pi$NNN problem. Our equations are formally
similar to the Yakubovskii--Grassberger--Sandhas equations for the
standard four--body problem. However, our equations couple the
four--body ($\pi$NNN) sector to the underlying three--nucleon system,
thereby allowing a consistent evaluation of pion production/absorption
amplitudes.


\begin{references}
\bibitem{note1} We remind that approaches of the
Thomas--Rinat--Avishai--Mizutani--Afnan--Blankleider type are introduced
through truncation of the underlying field theory. Therefore, also our
results are in fact ``exact" only if we accept such field theoretic
truncation.
\bibitem{note2} We briefly remind that $t_a$ represents either the NN
or $\pi$N (non polar) t--matrix in the 4--body sector, ${\cal V}$
is the sum over all the NN interactions in the 3--nucleon sector
(with explicit one--pion--exchange treatment), and in the two
free propagators for the 4--body and 3--nucleon sectors, $G_0$
and $g_0$ respectively, the nucleon dressings is treated differently
(see Refs. \cite{afnan80,avishai83,avishai83',cattapan93}).
\end{references}
\end{document}